\AtBeginDocument{%
  }

\documentclass[sigconf]{acmart}

\copyrightyear{2023}
\acmYear{2023}
\setcopyright{acmlicensed}
\acmConference[MM '23] {Proceedings of the 31st ACM International Conference on Multimedia}{October 29--November 3, 2023}{Ottawa, ON, Canada.}
\acmBooktitle{Proceedings of the 31st ACM International Conference on Multimedia (MM '23), October 29--November 3, 2023, Ottawa, ON, Canada}
\acmPrice{15.00}
\acmISBN{979-8-4007-0108-5/23/10}
\acmDOI{10.1145/3581783.3612396}

\usepackage{balance}

\usepackage{tikz}

\newtheorem{definition}{Definition}

\usepackage{multirow}

\acmSubmissionID{3034}

\begin{document}

\title{My Brother Helps Me: Node Injection Based Adversarial Attack on Social Bot Detection}


\author{Lanjun Wang}
\authornote{Both authors contributed equally to this research.}
\affiliation{%
  \institution{Tianjin University}
  \city{Tianjin}
  \country{China}
  }
\email{wanglanjun@tju.edu.cn}

\author{Xinran Qiao}
\affiliation{%
  \institution{Tianjin University}
  \city{Tianjin}
  \country{China}
  }
\authornotemark[1]
\email{qiaoxinran@tju.edu.cn}

\author{Yanwei Xie}
\affiliation{%
  \institution{Tianjin University}
  \city{Tianjin}
  \country{China}
  }
\email{xie\_yw@tju.edu.cn}

\author{Weizhi Nie}
\affiliation{%
  \institution{Tianjin University}
  \city{Tianjin}
  \country{China}
  }
\email{weizhinie@tju.edu.cn}

\author{Yongdong Zhang}
\affiliation{%
  \institution{University of Science and Technology of China}
  \city{Hefei}
  \country{China}
  }
\email{zhyd73@ustc.edu.cn}

\author{Anan Liu}
\authornote{Corresponding Author}
\affiliation{%
  \institution{Tianjin University}
  \city{Tianjin}
  \country{China}
  }
\email{liuanan@tju.edu.cn}

\renewcommand{\shortauthors}{Lanjun Wang et al.}


\begin{abstract}
  Social platforms such as Twitter are under siege from a multitude of fraudulent users. 
  In response, social bot detection tasks have been developed to identify such fake users.
  Due to the structure of social networks, the majority of methods are based on the graph neural network(GNN), which is susceptible to attacks.
  In this study, we propose a node injection-based adversarial attack method designed to deceive bot detection models. Notably, neither the target bot nor the newly injected bot can be detected when a new bot is added around the target bot. This attack operates in a black-box fashion, implying that any information related to the victim model remains unknown. 
  To our knowledge, this is the first study exploring the resilience of bot detection through graph node injection.
  Furthermore, we develop an attribute recovery module to revert the injected node embedding from the graph embedding space back to the original feature space, enabling the adversary to manipulate node perturbation effectively.
  We conduct adversarial attacks on four commonly used GNN structures for bot detection on two widely used datasets: Cresci-2015 and TwiBot-22. 
  The attack success rate is over 73\%  and the rate of newly injected nodes being detected as bots is below 13\% on these two datasets.
\end{abstract}

\ccsdesc[500]{Computing methodologies~Machine learning algorithms}
\ccsdesc[300]{Security and privacy~Human and societal aspects of security and privacy}

\keywords{Black-box Adversarial Attack, Bot Detection, Social Media}


\maketitle

\section{INTRODUCTION}  \label{sec:intro}

As social media becomes an increasingly integral part of people's daily lives, public opinion is now more heavily influenced by its content than ever before~\cite{tan2022efficient,chen2022and}. With millions of daily active users, Twitter is a thriving social media platform with significant influence in shaping public opinion. However, as well as bringing benefits, social media also poses a major threat~\cite{von2020helping}. The platform's immense power has led to the proliferation of a new type of users, known as social bots. These bots can amplify certain discussions at the expense of others and manipulate public opinion to achieve their own goals. Examples of such manipulation include extreme propaganda~\cite{berger2015isis}, promotion of political conspiracies~\cite{ferrara2020types,keller2020political}, and interference in elections~\cite{deb2019perils,ferrara2017disinformation}.

To address the various problems caused by social bots, social bot detection tasks have been developed. Existing social bot detection methods are usually divided into three categories: feature-based methods, text-based methods, and graph-based methods. Feature-based methods utilize user information for feature engineering and apply traditional classification algorithms to detect bots~\cite{kudugunta2018deep,feng2021satar}. Text-based methods use natural language processing techniques to process user tweets and user description texts for bot detection~\cite{cai2017detecting,knauth2019language}. Graph-based methods interpret the Twitter social network as a graph and use the concepts of network science and geometric deep learning for bot detection~\cite{dehghan2022detecting,pham2022bot2vec}.

Recent studies~\cite{feng2021botrgcn,feng2022twibot} have shown that graph-based methods achieve state-of-the-art performance in social bot detection. These methods can detect novel bots and overcome the various challenges associated with social bot detection. Graph-based methods typically employ Graph Neural Networks (GNNs) to detect bots by interpreting users as nodes and relationships as edges. Most GNNs follow a message passing scheme and achieve significant performance in many tasks~\cite{xu2020graph,rieck2019persistent,cao2020popularity} by iteratively aggregating representations of representation learning nodes from their neighbors. Despite their success, GNNs have been found to be highly vulnerable to adversarial attacks~\cite{xu2019topology,bojchevski2019adversarial,wang2020mgaattack,xu2020learning, sang2022benign}. Graph-based social bot detection methods rely on GNNs for processing social networks, making them similarly vulnerable to adversarial attacks. 

However, there is no attempt at social bot detection tasks because of the following difficulties.
Firstly, most of the existing adversarial attack methods are white-box attacks~\cite{hu2023hyperattack,castiglione2022scalable}, which require the attacker to master the victim model in advance. 
On the contrary, the information of the victim model is the key asset of the companies who manage social platforms, and thus it is infeasible for the attacker to access it.
Therefore, the practical adversarial attacks on social bot detection are in a black-box manner.
Secondly, too much modification on the social network can be noticed to lead to the failure of the attack, so it is necessary to maintain the imperceptibility of the attack method, which requires not perturbing too much information about the original network.  According to the studies on GNN adversarial attacks~\cite{tao2021single,dai2018adversarial,yuan2021meta,sun2020adversarial}, we choose to leverage the single node injection method~\cite{tao2021single} to control the change of the whole network on one node. Furthermore, since this newly injected node is also a fake user in the social network, there is a task-specific imperceptible requirement: the newly injected node cannot be detected by the victim model. This is a different constraint from the classical adversarial attacks on GNNs which control the number of the perturbed nodes~\cite{tao2021single} or the access graph range of the attackers~\cite{ma2020towards}.
Thirdly, most of the existing node injection adversarial attacks against GNNs are carried out in the intermediate embedding space~\cite{xu2020towards,tao2021single}, which leads to the attack generating an injected node in the form of embedding. However, for our specific adversarial attack on social bot detection, since the attacker needs to generate a new bot and inject it into the original social network to achieve the undetectable of the target bot, the original attributes of the injected bot have to be restored.

In this study, we address these challenges of black-box settings, imperceptibility, and attribute recovery in the adversarial attack on social bot detection.
First, we set up  a simple GNN structure based on embeddings from the original attribute space as a substitute model. This setting relies on the transferability of the adversarial samples to achieve the black-box attack.  
Then, a single-node injection adversarial attack approach, G-NIA~\cite{tao2021single}, is applied to address the imperceptibility of the attack from the perturbation aspect.
Moreover, to maintain the imperceptibility of the attack from the attribute aspect, we collect statistics on the attributes of human users, then convert them as a series of constraints, and apply these constraints to the newly injected node. 
Finally, we design an attribute recovery module to obtain the original features of the injected node from the embedding space.

In summary, this study has the following contributions:
\begin{itemize}
    \item In order to fool bot detection methods, we propose a black-box node injection-based adversarial attack method, which is to add a new bot around a target bot and to achieve both the target bot and the newly injected bot undetectable by the bot detection method.
    To the best of our knowledge, this is the first study on the node injection-based bot detection attack.
    \item We design an attribute recovery module to restore the node feature from the graph embedding space to the original feature space in order to make the adding node perturbation operable by the adversary.
    \item We attack four existing bot detection methods on two datasets (Cresci-2015 and TwiBot-22) to evaluate the generalizability and effectiveness of the attack models. The attack success rate is over 73\% and the rate of newly injected nodes being detected as bots is below 13\% on two datasets. Specifically, the newly injected node detection rate on Cresci-2015 is as low as 0.06\%. 
\end{itemize}

\section{RELATED WORK}

In this section, we introduce the vulnerability of bot detection and the adversarial attack on GNNs.

\subsection{Vulnerability of Bot Detection}

Existing bot detection methods are vulnerable to attack. The error of the bot detection model can be caused either by creating the bot scenarios or by carrying out the adversarial attack.

Torusdağ et al.~\cite{torusdaug2020we} investigate the vulnerability of existing social bot detection systems by creating their own bot scenarios instead of relying on public datasets. They experimentally show that the existing social bot detection model is unable to detect their social bots, thus demonstrating the vulnerability of these models. However, their study only serves to verify the vulnerability of the bot detection model and does not offer further discussion on potential solutions or improvements.
Kantartopoulos et al.~\cite{kantartopoulos2020exploring} obtain good results in adversarial attacks for bot detection by poisoning the training dataset. Their attack involved randomly modifying the labels of the bot nodes in the training set and copying new nodes based on the information of existing nodes. While this method is an adversarial attack for bot detection, it is not within the same scope as ours. This random attack method cannot specify nodes for attack and lacks flexibility. Additionally, this attack method directly uses the poisoned data to train the bot detection model, which can only be used to enhance the robustness of the bot detection model. In contrast, we propose an adversarial attack method based on node injection. Our method involves adding a new node and a new relationship to the original social network, which causes the bot detection model to generate classification errors for both the target bot node and the newly injected bot node. This approach greatly improves the accuracy and benefits of the attack.

\subsection{Adversarial Attack on GNN}

Extensive studies have shown that graph neural networks (GNNs) are vulnerable to various adversarial attacks~\cite{jia2020certified,ma2021graph,sun2020adversarial}. These attacks can perturb node attributes, graph structures, and labels~\cite{chen2020survey,xu2020adversarial}. For instance, Nettack~\cite{zugner2018adversarial} is a targeted attack that aims to deceive specific nodes by modifying their properties and the structure of the gradient bootstrap graph. Meta-attack~\cite{yang2021model} is a non-targeted attack based on meta-learning, but this method degrades the overall performance of GNNs. Additionally, G-NIA~\cite{tao2021single} is a targeted attack that adds only one new node and one new edge at a time, resulting in minimal disturbance to the original graph structure.

In this study, we use a similar approach to G-NIA to conduct an undetectable adversarial attack against the social bot detection model. G-NIA alone is insufficient to achieve our goal since this method can only generate the embeddings of injected nodes and requires reading the parameters of the victim model for a white-box attack. As we cannot access the specific information of the targeted bot detection model, we design a substitute model for a black-box attack. Additionally, to realize the attack on social networks, we need to obtain the attributes of injected nodes, including user descriptions, tweets, numerical metadata, and categorical metadata. To do so, we further propose an attribute recovery module to obtain the attributes of injected nodes from the generated embeddings.

\section{METHODOLOGY}

In this section, we delve into the specifics of our proposed method.  In Sec.~\ref{problem_definition}, we provide a definition of the problem we aim to address. Subsequently, in Sec.~\ref{3framework}, we present the overall framework of our methodology.
Finally, in Secs.~\ref{substitute_model}-\ref{attribute_recovery}, we specify the four primary modules in our framework.

\begin{figure*}[htp]
  \centering
  \includegraphics[width=\linewidth]{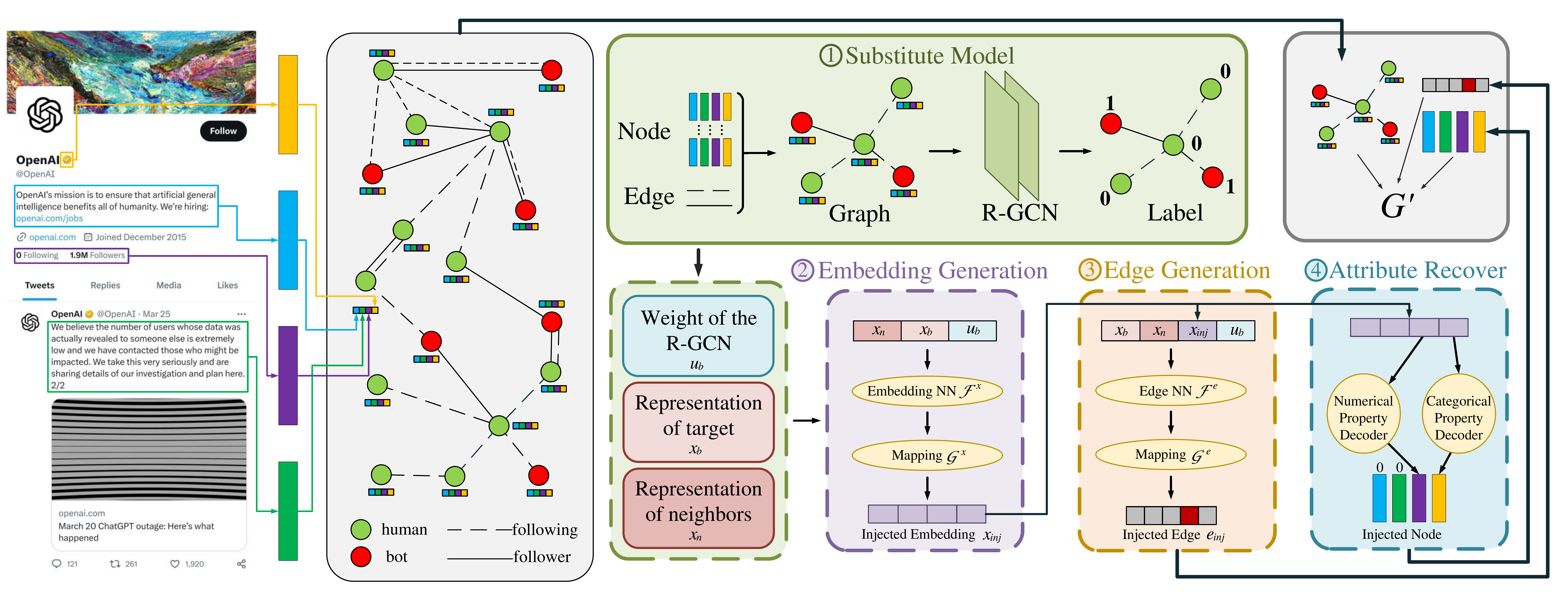}
  \caption{Framework of Single-node Adversarial Attack on Social
Bot Detection}
  \label{framework}
\end{figure*}

\subsection{Problem Definition}  \label{problem_definition}

\subsubsection{Objective}

Let $G = (V, E, A)$ represents a social network, where $V = \left\{1, 2, \dots, k \right\}$ constitutes the set of $k$ users, $E \subseteq V \times V$ defines the relationships of the users, and $A$ represents the attributes of users.

The goal of the social bot detection task is to achieve accurate prediction of user types in the social graph, i.e., $f(h|G)=y_h$ and $f(b|G)=y_b$, where $f(\cdot|G)$ denotes the detection outcome of the detection model with respect to the social network $G$, $h$ and $b$ represent human and bot respectively, $y_h$ and $y_b$ indicate the node type is human and bot respectively.

Based on the difficulties of the adversarial attack against the bot detection task mentioned in Sec.~\ref{sec:intro}, we adopt the single-node injection method to carry out the black-box attack. 
The problem is defined as follows: 
\begin{definition}[Single-node Adversarial Attack on Social Bot Detection]
Given a social network $G$, and a target bot $b_t$ which the attacker wants to evade the bot detection $f(\cdot)$, the single-node injection adversarial attack on bot detection is to obtain a new social network graph $G'$, where $G'$ is constituted by $G$ and one new bot $b_{inj}=(v_{inj}, e_{inj}, a_{inj})$, that is, $G'=(V \cup v_{inj}, E \cup e_{inj}, A \cup a_{inj})$, such that  $f(b_t|G')=y_h$ and  $f(b_{inj}|G')=y_h$.
\end{definition}
Specifically, this definition illustrates three requirements of this adversarial task: 1) the new social network graph $G'$ only has one new node (i.e., $v_{inj}$ with its attribute $a_{inj}$) and one new edge (i.e., $e_{inj}$) to connect the node to the original graph $G$, without altering any existing nodes or edges in $G$. This aims to achieve imperceptible perturbation on the social graph. 2) The original attribute $a_{inj}$  is required to be attained, which keeps the adversarial attack operable in practice. 3) Although the task is only to shield the target bot $b_t$, as the new injected node $b_{inj}$ is also a man-made bot, $b_{inj}$ is also required to be recognized as a human (i.e., $f(b_{inj}|G')=y_h$) to escape the bot detection.

\subsubsection{Threat Model}

As introduced in Sec.~\ref{sec:intro}, this study focuses on the black-box attack to keep it practical. That is, the adversary does not know the structure and weight information of the victim model because the victim model is the key asset of the companies who manage social networks. Furthermore, the adversary knows the target bot which they manage, as well as the corresponding social networks which are published and can be crawled from social media.

\subsection{Framework}  \label{3framework}

The overall framework of our method is shown in Figure~\ref{framework}, which has four major modules: \textcircled{1}substitute model, \textcircled{2}embedding generation, \textcircled{3}edge generation, and \textcircled{4}attribute recovery. In detail, as the adversarial attack is in a black-box manner, we train a Relational Graph Convolutional Network (R-GCN)~\cite{schlichtkrull2018modeling} as the \textcircled{1}substitute model, and leverage the transferability of the adversarial samples to fool some latest bot detectors. More details of victim models are in Sec.~\ref{experiment}.
Then, the node embeddings extracted from the \textcircled{1}substitute model as well as the weights of the \textcircled{1}substitute model are used to generate the embedding of the injected node  in \textcircled{2}embedding generation. By leveraging the embedding of the injected node as well as the information used in \textcircled{2},
\textcircled{3}edge generation obtains the injected edge. As illustrated in the definition, it is not enough to have the embedding of the new node, but the original attributes are required. 
Thus, \textcircled{4}attribute recovery restores the embedding of the injected node output by \textcircled{2}embedding generation into the original attributes. Finally, the new node can be created by the attacker and injected into the social graph to shield the target bot. 

\subsection{Substitute Model}   \label{substitute_model}

To launch an attack without knowing the specific information of the victim models, we devise a substitute model for transfer-based attacks.

For each user $v$ in the social graph $G$, the attribute $a$ includes four types, which are the description $D$, a set of tweets $T$, a series of numerical properties $N$, and a series of categorical properties $C$, that is $a=(D, T, N, C)$, where $D = \left\{ d_i \right\} ^L_{i=1}$ represents a user's description with $L$ words,  $T = \left\{t_i\right\}^M_{i=1}$ signify a user's $M$ tweets, $N = \left\{n_i\right\}^P_{i=1}$ denotes a user's numerical properties, and $C = \left\{c_i\right\}^Q_{i=1}$ indicate a user's categorical properties. 

For edges between users, given the following and being followed information, there are two types of edges: ``friend'' and ``follow''~\cite{cresci2015fame}, that is $R = \left\{ r_f, r_o \right\}$. 
Due to these different types of edges in the social network, the social network is a heterogeneous graph.
We represent the friend and follow neighborhoods of a user $v$ as $E_f(v)$ and $E_o(v)$, respectively, and all neighbors of $v$ is represented as $E_r(v) =  E_f(v)\cup E_o(v) $.

The substitute model consists of two parts: individual attribute encoding and structure-based feature transformation.
We mainly use four fully-connected layers  for individual attribute encoding and an R-GCN for structure-based feature transformation. Specifically, the fully-connected layer networks are used to encode the users' semantic information from their descriptions and tweets, as well as both user numerical and categorical property information, respectively, as illustrated in below (a)-(d).  Meanwhile, the R-GCN is to obtain the overall user representation by considering the heterogeneous social graph, which is specified as (e). 
In addition, we set the dimension of the total user embedding $D$. As the user embedding is concatenated by attribute embeddings as defined in Equation~(\ref{user_embedding}), the dimension of each attribute type is $\frac{D}{4}$, that is, the output dimension of the fully connected layer networks is  $\frac{D}{4}$.

The detailed processes are as follows.

\paragraph {a) User Description}

We utilize pre-trained RoBERTa~\cite{liu2019roberta} to encode user descriptions. Initially, we transform the words in the user description using RoBERTa:
\begin{equation}   \label{d1}
    \overline{d} = RoBERTa(\left\{ d_i \right\}^L_{i=1}), \quad \overline{d} \in \mathbb{R}^{D_s \times 1}
\end{equation}
where $\overline{d}$ represents the user description's representation, and $D_s$ corresponds to the embedding dimension of RoBERTa. Next, we extract representation vectors for the user's description:
\begin{equation}   \label{d2}
    x_d = \phi(W_D \cdot \overline{d} + b_D), \quad x_d \in \mathbb{R}^{\frac{D}{4} \times 1}
\end{equation}
where $W_D$ and $b_D$ are learnable parameters, $\phi$ is the activation function.

\paragraph{b) User Tweets}

We follow the same process as Equation~(\ref{d1}) and Equation~(\ref{d2}) on each tweet. Then, by averaging the representations of all tweets, we obtain the user tweet representation $x_t \in \mathbb{R}^{\frac{D}{4} \times 1}$.

\paragraph{c) User Numerical Properties}

We utilize numerical features that can be directly accessed through the Twitter API, as shown in Table~\ref{numerical_properties}. After performing z-score normalization, we obtain the representation of user numerical features $x_n \in \mathbb{R}^{\frac{D}{4} \times 1}$ using a fully connected layer.

\begin{table}[h]
  \caption{User numerical properties}
  \label{numerical_properties}
  \begin{tabular}{cl}
    \toprule
    Feature Name  &Description\\
    \midrule
    followers   & number of followers  \\
    active days   & number of active days  \\
    screen name length   & screen name character count  \\
    followings   & number of followings  \\
    status  & \small{number of public lists where this user belongs} 
\\
  \bottomrule
\end{tabular}
\end{table}

\paragraph{d) User Categorical Properties}

Similar to user numerical properties, we make use of directly available user categorical features from the Twitter API, which are displayed in Table~\ref{categorical_properties}. By employing one-hot encoding, concatenating, and transforming them with a fully-connected layer, we derive the representation for the user's categorical features $x_c \in \mathbb{R}^{\frac{D}{4} \times 1}$.

\begin{table}[h]
  \caption{User categorical properties}
  \label{categorical_properties}
  \begin{tabular}{cl}
    \toprule
    Feature Name  &Description\\
    \midrule
    protected   & protected or not  \\
    verified   & verified or not  \\
    default profile image   & the default profile image  \\
  \bottomrule
\end{tabular}
\end{table}

\paragraph{e) Overall User Embedding}

We encode the user description, tweets, numerical, and categorical properties, then concatenate them to serve as the user embedding. For each user $i\in V$, we represent the user embedding as:
\begin{equation}  \label{user_embedding}
    x_i = [{x_d}_i; {x_t}_i; {x_n}_i; {x_c}_i] \in \mathbb{R}^{D \times 1}
\end{equation}

{We employ an R-GCN~\cite{schlichtkrull2018modeling}} on the heterogeneous graph to learn user representations. The overall user embeddings in Equation~(\ref{user_embedding}) are used directly as the initially hidden vectors for nodes within the graph:
\begin{equation}   \label{rgcn_input}
    x_i^{(0)} = x_i, \quad x^{(0)}_i \in \mathbb{R}^{D \times 1}
\end{equation}

Then, we apply the $l$-th R-GCN layer:
\begin{equation}   \label{rgcn_transformation}
    x^{(l+1)}_i = \Theta_{self} \cdot x^{(l)}_i + \sum_{r \in R} \sum_{j \in E_r(i)} \frac{1}{|E_r(i)|} \Theta_r \cdot x^{(l)}_j
\end{equation}
where $\Theta$ is the projection matrix, $l=\{0,\ldots,L-1\}$, and $L$ is the number of edge types. We directly use the output of the R-GCN as the prediction label:
\begin{equation} \label{rgcn_output}
    \widehat{y}_i = x_i^{(L)} 
\end{equation}
{Here, we jointly train the attribute encoding and overall feature transformation in the substitute model.}
The loss function of the substitute model is constructed as follows:
\begin{equation}
    L_{s} = - \sum_{i \in V}[y_i log(\widehat{y}_i)+(1-y_i)log(1-(\widehat{y}_i)] + \lambda \sum_{w \in \theta_s} w^2
\end{equation}
where $y_i$ is the ground-truth label and $\theta_s$ are all parameters of the substitute model.

\subsection{Embedding Generation}  \label{embedding_generation}

We use the embeddings of the target node $x_{b_t}$ and its first-order neighbors $x_n$, to guide the generation of the injected node, where $x_n$ is the average embedding of all first-order neighbors of the target node $b_t$.
Since feature transformation maps the embedding space to the label space {as shown in Equations~(\ref{rgcn_input})-(\ref{rgcn_output}}), we adopt the column of the feature transformation weights to represent the label class $u_{b_t}$. The process of constructing $u_{b_t}$ is as follows. Since there are two types of edges (i.e., ``friend'' and ``follow'') in the graph, the weight of the R-GCN is divided into $W_f$ and $W_o$ and then aggregated with a fully-connected layer $f_W$, that is $W=f_W(W_f, W_o)$. Finally, $u_{b_t} = [W[:,y_b]; W[:,y_h]; W]$, where $W[:,y_b]$ represents the column of $W$ with respect to the label of before attacking (i.e. bot, $y_b$), and $W[:,y_h]$ represents the column of $W$ with respect to the expected label of after attacking (i.e. human, $y_h$).

Using the representations described above, we utilize two fully-connected layers $\mathcal{F}^x$ and a Multi-Layer Perceptron (MLP) $\mathcal{G}^x$ to generate the embedding of the injected node, denoted as $x_{inj}$:
\begin{equation}   \label{eq:emb_gen}
\begin{aligned}
& x_{inj}=\mathcal{G}^x(\mathcal{F}^x(x_{b_t}, x_n, u_{b_t},G;\theta ^ *_x))\\
 \mathcal{F}^x(x_{b_t}, x_n, & u_{b_t},G;\theta^*_x) = \sigma([x_n;x_{b_t};u_{b_t}]W^x_0+b^x_0)W^x_1+b^x_1
\end{aligned}
\end{equation}
where $\theta^*_x = \left\{W^x_0, b^x_0, W^x_1, b^x_1\right\}$ are trainable weights. The mapping function $\mathcal{G}^x$ maps the output of $\mathcal{F}^x$ to the designated embedding space of the original graph, making the embeddings similar to existing nodes.

\subsection{Edge Generation}   \label{edge_generation}

The injected edges serve to spread the attributes of the injected node to the target nodes. The injected edge is limited to the target node and its first-order neighbors, meaning that the injected node must be at least a second-order neighbor of the target node.
To capture the coupling effect between network structure and node features, we jointly model the injected embeddings and edges. 

Specifically, we use the injected embeddings to guide the generation of injected edges.
To guide the generation of the injected edge $e_{inj}$, we include the generated injected embedding $x_{inj}$ along with the information of the target bot and its neighbors as used in Equation~(\ref{eq:emb_gen}). We employ two fully-connected layers $\mathcal{F}^e$ and an {MLP} $\mathcal{G}^e$ to generate $e_{inj}$ as follows:
\begin{equation}   \label{eq:edge_gen}
\begin{aligned}
& e_{inj}=\mathcal{G}^e(\mathcal{F}^e(x_{inj},x_{b_t}, x_n, u_{b_t},G;\theta ^ *_e))\\
\mathcal{F}^e(x_{inj},x_{b_t}, & x_n, u_{b_t},G;\theta^*_e) = \sigma([x_{b_t};x_n;x_{inj};u_{b_t}]W^e_0+b^e_0)W^e_1+b^e_1
\end{aligned}
\end{equation}
where $\theta^*_e = \left\{W^e_0, b^e_0, W^e_1, b^e_1\right\}$ are trainable weights.

Once the injected embedding and edge are generated, we inject the injected node into the original graph $G$ to obtain the perturbed graph $G'$. We then feed $G'$ into the substitute model and compute the attack loss as follows:
\begin{equation}
    L_{atk}=\sum_{b_t \in V_b}(S'_{b_t,y_b}-S'_{b_t,y_h})
\end{equation}
where $S'_{b_t,y}$ denotes the predicted label probability of the substitute model in the new social network graph $G'$ with respect to the target bot $b_t$ on the corresponding label (i.e. $y_h$ and $y_b$).
Here we conduct joint training on the embedding generation module and edge generation module.
The optimization process aims to minimize the attack loss $L_{atk}$, which guides the training process. We employ gradient descent to iteratively optimize $L_{atk}$ until convergence.

\subsection{Attribute Recovery}   \label{attribute_recovery}

The social network bot detection model detects users by extracting features from the raw user data. However, the injected node embeddings generated in Sec.~\ref{embedding_generation} represent the total user features that have been extracted by the feature extractor in the substitute model, which is not consistent with the bot detection model whose input is the original user features. To address this, we propose a module named attribute recovery to recover the generated injected node embeddings.

\subsubsection{User Description \& User Tweets}

It is challenging to recover text features, so we directly set the text features for the user description and user tweets to 0 for the injected node. This means that the injected node does not contain any text information.

\subsubsection{User Numerical Properties}

To recover the numerical features of the injected node, we train a multi-layer perceptron (MLP) using the preprocessed numerical data $N = \left\{n_i\right\}^P_{i=1}$ and the numerical embeddings $x_n$ extracted by the feature extractor of the substitute model.
First, we calculate the loss per user at the original feature level:
\begin{equation}   \label{num_mlp_loss1}
    l_{n1} = \sum_{i=1}^{P} \mid MLP_n(x_{n_i}) - n_{i} \mid
\end{equation}
Next, we also constrain the effects of the numerical MLP at the feature level:
\begin{equation}
    l_{n2} = \sum_{i=1}^{P} \mid \phi (W_N \cdot MLP_n(x_{n_i}) + b_N )- x_{n_{i}} \mid
\end{equation}
Finally, we add the losses for each user $j$ as a loss function of the numerical MLP:
\begin{equation}   \label{num_loss}
    L_n = \sum_{j \in V} (l_{n1} + \alpha l_{n2} )
\end{equation}
where $\alpha$ is a constant, and here we set $\alpha = 0.01$.

To ensure that the recovered numerical features are realistic, we reverse the z-score normalization on the recovered features to obtain the original integer numerical features. Then, we constrain the numerical features to their respective appropriate ranges and perform z-score normalization again to obtain the final user numerical properties. The constraints are carefully designed based on the statistics of the corresponding dataset where the node is to be injected, in order to 1) achieve successful attacks to shield the target bot, 2) make the injected node imperceptible, as well as 3) the injection process operable by the attacker. Detailed setting criteria can be referred to in Sec.~\ref{constraint_settings}.

\subsubsection{User Categorical Properties}

Similar to the attribute recovery of the user numerical properties, we train another MLP using the data $C = \left\{c_i\right\}^Q_{i=1}$ obtained after data preprocessing and the categorical embeddings $x_c$ processed by the feature extractor of the substitute model.
{For the categorical MLP, we calculate each user's losses at the original feature level only, similar to Equation~(\ref{num_mlp_loss1}), and then sum all users' losses as the loss function of categorical property recovery}.
We recover the user categorical properties to the one-hot encoding by adding a constraint during the attribute recovery process.

\subsubsection{Edge Type}

After generating the injected node with the injected embedding and recovering its attributes, we need to connect it to the target node in the subgraph. Since the injected node is a bot, we set the injected edge as "follow" to make it easier to establish.

\section{EXPERIMENT}   \label{experiment}

In this section, we conduct experiments to show the efficacy of our method.
In Sec.~\ref{experiment_settings}, we introduce the experiment settings. Then, Sec.~\ref{overall_performance} shows the overall performance of our method. On this basis, we conduct ablation experiments, parameter analysis, and transferability analysis in Sec.~\ref{ablation_study}, Sec.~\ref{parameter_analyse}, and Sec.~\ref{transferability_analyse}, respectively. 

\subsection{Experiment Settings}  \label{experiment_settings}
\subsubsection{Datasets}
We conduct experiments to evaluate the effectiveness of our method on two datasets: Cresci-2015~\cite{cresci2015fame} and TwiBot-22~\cite{feng2022twibot}. These datasets comprise heterogeneous graph data from social networks, and their statistics are shown in Table~\ref{statistics}.

\begin{table}[h]
  \caption{Statistics of the datasets}
  \label{statistics}
  \begin{tabular}{cccccc}
    \toprule
    Dataset      & Human   & Bot   & Tweet  & Edge   \\
    \midrule
    Cresci-2015   & 1,950   & 3,351   & 2,827,757   & 7,086,134  \\
    TwiBot-22   & 860,057   & 139,943   & 86,764,167   & 170,185,937  \\
  \bottomrule
\end{tabular}
\end{table}

Due to NVIDIA RTX 3090Ti GPU memory limitations, we are unable to convert the complete social network graph provided by the TwiBot-22 dataset into a sparse matrix for quickly obtaining the first-order neighbors of the target as required in Equation~\ref{eq:emb_gen} and Equation~\ref{eq:edge_gen}.
Therefore, We use a community selection algorithm to select about 50,000 nodes to form five subgraphs for experiments. The statistics of the subgraphs are shown in Table~\ref{sub_graph}.

\begin{table}[h]
  \caption{Statistics of the five subgraphs selected on TwiBot-22}
  \label{sub_graph}
  \begin{tabular}{cccc}
    \toprule
    Graph      & Human  & Bot  & Edge    \\
    \midrule
    subgraph 1   & \multicolumn{1}{r}{45,081}   & \multicolumn{1}{r}{4,491}   & \multicolumn{1}{r}{993,930} \\
    subgraph 2   & \multicolumn{1}{r}{44,864}   & \multicolumn{1}{r}{4,276}   & \multicolumn{1}{r}{950,365} \\
    subgraph 3   & \multicolumn{1}{r}{46,079}   & \multicolumn{1}{r}{4,278}   & \multicolumn{1}{r}{953,733} \\
    subgraph 4   & \multicolumn{1}{r}{46,267}   & \multicolumn{1}{r}{4,542}   & \multicolumn{1}{r}{963,507} \\
    subgraph 5   & \multicolumn{1}{r}{45,553}   & \multicolumn{1}{r}{4,496}   & \multicolumn{1}{r}{983,006} \\
    whole graph   & \multicolumn{1}{r}{860,057}   & \multicolumn{1}{r}{139,943}   & \multicolumn{1}{r}{170,185,937} \\
  \bottomrule
\end{tabular}
\end{table}

\subsubsection{Constraint Settings}   \label{constraint_settings}

To make the injected node more realistic, we set constraints to adjust its attributes as introduced in Sec.~\ref{attribute_recovery}. The constraints for injected nodes in the two datasets are shown in Table~\ref{constraint_15} and Table~\ref{constraint_22}.

In order to make the user numerical properties of the injected node realistic and easy to implement, we apply the following constraint-setting criteria.
\begin{itemize}
    \item The number of followers is set to 0, which indicates that the injected node does not need to be noticed by any others. That is to say, this indicates that the attacker does not need to buy any fan for this injected node.
    \item The maximum number of active days is set to 100, which indicates that the attacker does not need too much time to prepare for the shield. Noted this number is far lower than the average active days in the social networks.
    \item The screen name length and the number of followings to the maximum allowed by Twitter, as these are easily attainable attributes.
    \item The numbers of statuses are on the same orders of magnitudes as most users in the corresponding original social networks.
\end{itemize}

\begin{table}[h]
\caption{Mean and standard deviation of the numerical features as well as the constraints on Cresci-2015}
\label{constraint_15}
\begin{tabular}{cccc}
\toprule
Feature Name           & Mean                                  & Std                                   & Constraints                           \\ \hline
followers count        &  246  & 5,879   & 0  \\
active days        &  3,201  & 461   & 100 \\
screen name length & 11  & 3  & 15  \\
following count      &  386  & 561  & 5,000  \\
status      & 3  & 25 & 500 \\
\bottomrule
\end{tabular}
\end{table}

\begin{table}[h]
\caption{Mean and standard deviation of the numerical features as well as the constraints on TwiBot-22}
\label{constraint_22}
\begin{tabular}{cccc}
\toprule
     Feature Name      & Mean                                  & Std                                   & Constraints            \\ \hline
followers count        &  41,230  & 602,078   & 0 \\
active days        &  2,128  & 1,633  & 100  \\
screen name length & 14   & 7  & 15   \\
following count      &  2,251  & 15,782  & 5,000   \\
status      & 21,689  & 120,247  & 40,000 \\
\bottomrule
\end{tabular}
\end{table}

\subsubsection{Victim Models}

We evaluate the effectiveness of our method by attacking four GNNs that are used for bot detection, 
which are benchmarks given by the Twibot-22~\cite{feng2022twibot} dataset. The four methods use Graph Convolutional Network (GCN)~\cite{kipf2016semi}, Heterogeneous Graph Transformer (HGT)~\cite{hu2020heterogeneous}, Simple Heterogeneous Graph Neural Network (Simple-HGN)~\cite{lv2021we}, and Relational Graph Convolutional Network (R-GCN)~\cite{schlichtkrull2018modeling} respectively for social bot detection. These four methods leverage text information from user descriptions and tweets, user numerical and categorical property information, as well as heterogeneous graphs based on user relationships to learn user representations for bot detection tasks on Twitter. Among them, the method based on R-GCN~\cite{feng2021botrgcn} is different from our substitute model. It jointly encodes the various original attributes of each user and inputs them to two R-GCNs. Then, it uses a fully-connected layer to get the prediction label. 
{Meanwhile, as introduced in Sec.~\ref{substitute_model}, we encode the four types of original attributes respectively, splice them together, and then input them to a network with one R-GCN to  predict results directly.}

\subsubsection{Evaluation Metrics}

To demonstrate the effectiveness of our method, we set two evaluation metrics as follows:

\paragraph{Attack success rate}

We measure the success of the attack by evaluating whether the victim model can detect the target bot to the new social network graph $G'$. A successful attack occurs when the target bot node is classified as a human. That is to say, the higher the attack success rate, the better effectiveness of the attack.  
\paragraph{New node detected as bot}

We evaluate the imperceptibility of our injected node by measuring whether the victim model can detect it as a bot. A successful attack occurs when the injected node is classified as a human. That is to say, the lower rate of the new node detected as a bot, the better imperceptibility of the attack.  

\subsubsection{Implementation Details}

When training the substitute model, the learning rate is set as 1e-2, and the epoch is set as 150. 
In the training for embedding generation and edge generation, we set a maximum of 500 epochs. If the misclassification rate of the validation set does not increase for 5 consecutive epochs, the training is stopped. Here, we set the batch size to 32 and the learning rate to 1e-5.
For the attribute recovery, we set the initial learning rate as 1e-2, and the learning rate drops to 1e-5 with  training.
For the numerical property recovery loss in Equation~\ref{num_loss}, we set $ \alpha = 0.01$.
Each experiment is repeated five times to ensure reliability.

\subsection{Overall Performance}   \label{overall_performance}

We evaluate the effectiveness of our method on two datasets, Cresci-2015 and TwiBot-22, and the results are presented in Table~\ref{overall_15} and Table~\ref{overall_22}, respectively. 

\begin{table}[h]
  \caption{Overall performance on Cresci-2015 (\%)}
  \label{overall_15}
  \begin{tabular}{ccc}
    \toprule
    Method      & attack success rate  & new node become bot  \\
    \midrule
    GCN   & 95.68 ± 1.44   & 0.00 ± 0.00  \\
    HGT   & 94.79 ± 1.18   & 0.06 ± 0.12  \\
    Simple-HGN   & 95.74 ± 1.25   & 0.00 ± 0.00   \\
    R-GCN     & 95.74 ± 1.50   & 0.06 ± 0.12   \\
  \bottomrule
\end{tabular}
\end{table}

\begin{table}[h]
  \caption{Overall performance on TwiBot-22 (\%)}
  \label{overall_22}
  \begin{tabular}{ccc}
    \toprule
    Method      & attack success rate  & new node become bot \\
    \midrule
    GCN  & 93.97 ± 5.43   & 2.66 ± 5.09  \\
    HGT   & 89.37 ± 3.56   & 5.40 ± 10.80  \\
    Simple-HGN   & 74.94 ± 2.16   & 7.39 ± 14.78   \\
    R-GCN     & 73.73 ± 1.71   & 12.94 ± 19.19   \\
  \bottomrule
\end{tabular}
\end{table}

Our method successfully achieves significant attack results on both datasets.
However, it is noteworthy that the attack results on Cresci-2015 are better than those on TwiBot-22. This can be attributed to the increasing complexity of social networks and the advancements in camouflage technologies employed by bot users.

\subsection{Ablation Study}   \label{ablation_study}

To demonstrate the validity of our adversarial attack method, we conduct ablation experiments on our method. Since the substitute model and attribute recovery module are essential for data conversion, they cannot be removed. Therefore, we  perform ablation on the embedding generation and edge generation modules. We choose the TwiBot-22 dataset, which is more representative of the current social network environment.

In the ablation experiment for embedding generation, we directly assign the embedding of the target node to the injected node to explore the importance of the embedding generation module in adversarial attacks. The experimental results are shown in Table~\ref{ablation_embedding}. It is obvious that the attack effect decreases after removing the embedding generation module. This shows that the embedding generation module plays a vital role in both the attack success rate and the probability of the new node becoming a bot.

\begin{table}[h]
\caption{Ablation study for embedding generation on TwiBot-22 (\%)}
 \label{ablation_embedding}
\footnotesize
\begin{tabular}{ccccc}
\toprule
\multirow{2}{*}{Method}        & \multicolumn{2}{c}{attack success rate}                                           & \multicolumn{2}{c}{new node become bot}                                           \\
           & ours                                  & assign                                   & ours                                  & assign                                   \\ \hline
GCN        & \textbf{93.97 ± 5.43}   & 85.99 ± 2.75   & \textbf{2.66 ± 5.09}   & 39.24 ± 1.32  \\
HGT        & \textbf{89.37 ± 3.56}   & 84.36 ± 2.66   & \textbf{5.40 ± 10.80}   & 26.86 ± 1.69  \\
Simple-HGN & \textbf{74.94 ± 2.16}   & 74.10 ± 1.19   & \textbf{7.39 ± 14.78}   & 37.54 ± 1.15  \\
R-GCN      & 73.73 ± 1.71   & \textbf{74.91 ± 2.17}    & \textbf{12.94 ± 19.19}   & 35.29 ± 1.82  \\
\bottomrule
\end{tabular}
\end{table}

In the ablation experiment for edge generation, we randomly select a node in the first-order subgraph of the target node to connect with the injected node. The experimental results are shown in Table~\ref{ablation_edge}.
Overall, our approach achieves better performance, but the improvement is modest. This may be due to the failure to select all first-order nodes around the target bot during dataset construction and subgraph selection, resulting in a limited selection range of edge generation. We consider that in practical application if all the first-order neighbors can be selected to construct the subgraph, our method can achieve better results.

\begin{table}[h]
\caption{Ablation study for edge generation on TwiBot-22 (\%)}
 \label{ablation_edge}
\footnotesize
\begin{tabular}{ccccc}
\toprule
\multirow{2}{*}{Method}           & \multicolumn{2}{c}{attack success rate}                                           & \multicolumn{2}{c}{new node become bot}                                           \\
           & ours                                  & random                                  & ours                                  & random                                  \\ \hline
GCN        & \textbf{94.51 ± 1.57}   & 93.80 ± 2.19   & 2.66 ± 5.09   & \textbf{2.17 ± 4.12}  \\
HGT        & \textbf{89.37 ± 3.56}   & 86.48 ± 1.02   & 5.40 ± 10.80   & \textbf{4.34 ± 8.67}  \\
Simple-HGN & 74.94 ± 2.16   & \textbf{75.52 ± 1.15}   & 7.39 ± 14.78   & \textbf{6.37 ± 12.74}  \\
R-GCN      & 73.73 ± 1.71   & \textbf{74.39 ± 1.46}   & 12.94 ± 19.19   & \textbf{8.37 ± 10.83}  \\
\bottomrule
\end{tabular}
\end{table}

\begin{table*}[h]
\caption{The number of R-GCNs in the substitute model on Cresci-2015 (\%)}
\label{rgcn_num_15}
\footnotesize
\begin{tabular}{ccccccc}
\toprule
\multirow{2}{*}{Method}            & \multicolumn{3}{c}{attack success rate}                                           & \multicolumn{3}{c}{new node become bot}                                            \\
          & 1  & 2  & 3  & 1  & 2  & 3  \\ \hline
GCN        &  95.68 ± 1.44  & \textbf{95.86 ± 1.66}   & 95.62 ± 1.60   & \textbf{0.00 ± 0.00}   & \textbf{0.00 ± 0.00}   &0.12 ± 0.15  \\
HGT        &  94.79 ± 1.18   &87.40 ± 10.19   &\textbf{95.66 ± 2.19}   &\textbf{0.06 ± 0.12}   &28.23 ± 37.35   &13.55 ± 12.16 \\
Simple-HGN &  95.74 ± 1.25   &\textbf{95.80 ± 1.74}   &94.57 ± 1.59   &\textbf{0.00 ± 0.00}   &28.64 ± 38.83   &1.66 ± 1.41 \\
R-GCN      &  \textbf{95.74 ± 1.50}   &95.62 ± 1.39   &95.62 ± 1.59   &\textbf{0.06 ± 0.12}   &3.37 ± 6.74   &1.66 ± 1.07 \\
\bottomrule
\end{tabular}
\end{table*}

\begin{table*}[h]
\caption{The number of R-GCNs in the substitute model on TwiBot-22 (\%)}
 \label{rgcn_num_22}
\footnotesize
\begin{tabular}{ccccccc}
\toprule
\multirow{2}{*}{Method}            & \multicolumn{3}{c}{attack success rate}                                           & \multicolumn{3}{c}{new node become bot}                                      \\
          & 1  & 2  & 3  & 1  & 2  & 3  \\ \hline
GCN        & 93.97 ± 5.43   & 95.08 ± 2.79   & \textbf{95.60 ± 1.50}   & 2.66 ± 5.09   & 4.82 ± 7.27    & \textbf{0.40 ± 0.49} \\
HGT        & \textbf{89.37 ± 3.56}   & 85.67 ± 6.46   & 86.16 ± 5.76   & 5.40 ± 10.80   & 3.04 ± 4.60   & \textbf{0.04 ± 0.09}  \\
Simple-HGN & \textbf{74.94 ± 2.16}   & 72.07 ± 2.97    & 72.76 ± 5.05  & 7.39 ± 14.78   & 1.25 ± 1.63   & \textbf{0.00 ± 0.00}  \\
R-GCN      & 73.73 ± 1.71   & \textbf{79.40 ± 5.05}   & 78.83 ± 1.42   & 12.94 ± 19.19   & 3.20 ± 2.74  & \textbf{0.04 ± 0.09}   \\
\bottomrule
\end{tabular}
\end{table*}

\subsection{Parameter Analyse}    \label{parameter_analyse}

\subsubsection{Substitute model}

In this experiment, we investigate the impact of the substitute model on the performance of the attack. 
Besides R-GCN used in the overall performance as shown in Sec.~\ref{substitute_model}, we implement the framework based on a GCN substitute model.
We use a GCN layer to directly replace the R-GCN in the substitute model. Since GCN can only handle homogeneous graphs, we do not add edge-type information during input.
The outputs of GCN are consistent with the output of R-GCN, which are predictive labels.

\begin{table}[h]
\caption{GNN types in substitute model on Cresci-2015 (\%)}
\label{gnn_15}
\footnotesize
\begin{tabular}{ccccc}
\toprule
\multirow{2}{*}{Method}             & \multicolumn{2}{c}{attack success rate}                                           & \multicolumn{2}{c}{new node become bot}                                           \\
           & R-GCN                                  & GCN                                   & R-GCN                                  & GCN                                   \\ \hline
GCN        &  95.68 ± 1.44  & \textbf{96.51 ± 1.42}   & \textbf{0.00 ± 0.00}   & \textbf{0.00 ± 0.00}  \\
HGT        &  \textbf{94.79 ± 1.18}   &94.02 ± 0.51    & \textbf{0.06 ± 0.12}   & 8.46 ± 13.29 \\
Simple-HGN &  95.74 ± 1.25   & \textbf{96.75 ± 1.24}   &\textbf{0.00 ± 0.00}   & 11.12 ± 22.25 \\
R-GCN      &  95.74 ± 1.50  & \textbf{96.16 ± 0.82}  & \textbf{0.06 ± 0.12}   & 11.54 ± 23.08 \\
\bottomrule
\end{tabular}
\end{table}

\begin{table}[h]
\caption{GNN types in substitute model on TwiBot-22 (\%)}
\label{gnn_22}
\footnotesize
\begin{tabular}{ccccc}
\toprule
\multirow{2}{*}{Method}             & \multicolumn{2}{c}{attack success rate}                                           & \multicolumn{2}{c}{new node become bot}                                    \\
           & R-GCN                                  & GCN                                   & R-GCN                                  & GCN                                   \\ \hline
GCN        & \textbf{93.97 ± 5.43}   & 80.22 ± 10.87   & \textbf{2.66 ± 5.09}   & 66.79 ± 36.76  \\
HGT        & \textbf{89.37 ± 3.56}   & 84.13 ± 5.22   & \textbf{5.40 ± 10.80}   & 75.91 ± 32.64  \\
Simple-HGN & 74.94 ± 2.16   & \textbf{76.89 ± 1.96}   & \textbf{7.39 ± 14.78}   & 79.74 ± 32.17  \\
R-GCN      & 73.73 ± 1.71   & \textbf{74.61 ± 2.98}   & \textbf{12.94 ± 19.19}   & 71.93 ± 35.65  \\
\bottomrule
\end{tabular}
\end{table}

As shown in Table~\ref{gnn_15} and Table~\ref{gnn_22}, we can observe that the substitute model has a significant impact on the adversarial attack effect. Specifically, R-GCN achieves the highest overall attack success rate, and the probability of the injected node being detected as a bot is lower, especially on the TwiBot-22 dataset. This may be because R-GCN incorporates edge information when processing information, resulting in the weight of R-GCN containing more informative features than GCN.

\subsubsection{Number of R-GCNs in the Substitute Model}

In this experiment, we investigate the effect of the number of R-GCNs in the substitute model on the performance of the adversarial attack method, and the results are presented in Table~\ref{rgcn_num_15} and Table~\ref{rgcn_num_22}. 

The experimental results show that the number of R-GCNs in the substitute model has little effect on the adversarial attack effect. Therefore, we choose only one R-GCN as the structure of feature transformation in the substitute model, which requires the least amount of computation while carrying out the effective attack.

\subsection{Transferability Analyse}   \label{transferability_analyse} 

Sec.~\ref{overall_performance} demonstrates the transferability between models because we train on the substitute model and test on victim models.  
In this experiment, we investigate the transferability between data, namely whether the attack model can be trained on a subset of social networks and used to attack nodes on the complete social network. Since Twibot-22 has been divided into 5 subgraphs as shown in Table~\ref{sub_graph}, we train the attack model on one subgraph to attack bot nodes on others. The results are shown in Table~\ref{transferability}.

\begin{table}[h]
\caption{Transferability analysis on TwiBot-22 (\%)}
\footnotesize
\label{transferability}
\begin{tabular}{ccccc}
\toprule
\multirow{2}{*}{Method}             & \multicolumn{2}{c}{attack success rate}                                           & \multicolumn{2}{c}{new node become bot}                                          \\
           & same    & others   & same    & others                                   \\ 
           \hline
GCN        & 93.97 ± 5.43   & \textbf{94.03 ± 3.49}   & \textbf{2.66 ± 5.09}   & 7.72 ± 20.99  \\
HGT        & \textbf{89.37 ± 3.56}   & 88.29 ± 5.19   & \textbf{5.40 ± 10.80}   & 8.27 ± 24.09  \\
Simple-HGN & \textbf{74.94 ± 2.16}   & 74.92 ± 1.88   & \textbf{7.39 ± 14.78}   & 8.44 ± 24.06  \\
R-GCN      & 73.73 ± 1.71   & \textbf{74.09 ± 2.19}   & 12.94 ± 19.19   & \textbf{16.65 ± 20.46}  \\
\bottomrule
\end{tabular}
\end{table}

The experimental results demonstrate that our adversarial attack method exhibits good transferability.
By training on one social network, it can achieve good attack performance on different social networks. This shows that our model is general and not limited to a specific social network.

\section{CONCLUSION}

In this study, we propose an adversarial attack framework to hide bot users in social networks.
We employ the single-node injection method to conduct a black-box attack, and subsequently perform attribute recovery on the injected node embedding. Our attack makes neither the target bot node nor the injected bot node detected as a bot by the victim model.
Experimental results on the Cresci-2015 and TwiBot-22 datasets demonstrate the effectiveness of our method in achieving a high attack success rate and making the injected node undetectable.

\begin{acks}
This work is supported in part by the National Natural Science Foundation of China (U21B2024, 62202329).
\end{acks}

\bibliographystyle{ACM-Reference-Format}
\balance
\bibliography{sample-base}


\begin{thebibliography}{46}


\ifx \showCODEN    \undefined \def \showCODEN     #1{\unskip}     \fi
\ifx \showDOI      \undefined \def \showDOI       #1{#1}\fi
\ifx \showISBNx    \undefined \def \showISBNx     #1{\unskip}     \fi
\ifx \showISBNxiii \undefined \def \showISBNxiii  #1{\unskip}     \fi
\ifx \showISSN     \undefined \def \showISSN      #1{\unskip}     \fi
\ifx \showLCCN     \undefined \def \showLCCN      #1{\unskip}     \fi
\ifx \shownote     \undefined \def \shownote      #1{#1}          \fi
\ifx \showarticletitle \undefined \def \showarticletitle #1{#1}   \fi
\ifx \showURL      \undefined \def \showURL       {\relax}        \fi
\providecommand\bibfield[2]{#2}
\providecommand\bibinfo[2]{#2}
\providecommand\natexlab[1]{#1}
\providecommand\showeprint[2][]{arXiv:#2}

\bibitem[Berger and Morgan(2015)]%
        {berger2015isis}
\bibfield{author}{\bibinfo{person}{Jonathon~M Berger} {and}
  \bibinfo{person}{Jonathon Morgan}.} \bibinfo{year}{2015}\natexlab{}.
\newblock \showarticletitle{The ISIS Twitter Census: Defining and describing
  the population of ISIS supporters on Twitter}.
\newblock  (\bibinfo{year}{2015}).
\newblock


\bibitem[Bojchevski and G{\"{u}}nnemann(2019)]%
        {bojchevski2019adversarial}
\bibfield{author}{\bibinfo{person}{Aleksandar Bojchevski} {and}
  \bibinfo{person}{Stephan G{\"{u}}nnemann}.} \bibinfo{year}{2019}\natexlab{}.
\newblock \showarticletitle{Adversarial Attacks on Node Embeddings via Graph
  Poisoning}. In \bibinfo{booktitle}{\emph{ICML}}, Vol.~\bibinfo{volume}{97}.
  \bibinfo{pages}{695--704}.
\newblock


\bibitem[Cai et~al\mbox{.}(2017)]%
        {cai2017detecting}
\bibfield{author}{\bibinfo{person}{Chiyu Cai}, \bibinfo{person}{Linjing Li},
  {and} \bibinfo{person}{Daniel Zeng}.} \bibinfo{year}{2017}\natexlab{}.
\newblock \showarticletitle{Detecting Social Bots by Jointly Modeling Deep
  Behavior and Content Information}. In \bibinfo{booktitle}{\emph{CIKM}}.
  \bibinfo{pages}{1995--1998}.
\newblock


\bibitem[Cao et~al\mbox{.}(2020)]%
        {cao2020popularity}
\bibfield{author}{\bibinfo{person}{Qi Cao}, \bibinfo{person}{Huawei Shen},
  \bibinfo{person}{Jinhua Gao}, \bibinfo{person}{Bingzheng Wei}, {and}
  \bibinfo{person}{Xueqi Cheng}.} \bibinfo{year}{2020}\natexlab{}.
\newblock \showarticletitle{Popularity Prediction on Social Platforms with
  Coupled Graph Neural Networks}. In \bibinfo{booktitle}{\emph{WSDM}}.
  \bibinfo{pages}{70--78}.
\newblock


\bibitem[Castiglione et~al\mbox{.}(2022)]%
        {castiglione2022scalable}
\bibfield{author}{\bibinfo{person}{Giuseppe Castiglione},
  \bibinfo{person}{Gavin Ding}, \bibinfo{person}{Masoud Hashemi},
  \bibinfo{person}{Christopher Srinivasa}, {and} \bibinfo{person}{Ga Wu}.}
  \bibinfo{year}{2022}\natexlab{}.
\newblock \showarticletitle{Scalable Whitebox Attacks on Tree-based Models}.
\newblock \bibinfo{journal}{\emph{CoRR}} (\bibinfo{year}{2022}).
\newblock


\bibitem[Chen et~al\mbox{.}(2020)]%
        {chen2020survey}
\bibfield{author}{\bibinfo{person}{Liang Chen}, \bibinfo{person}{Jintang Li},
  \bibinfo{person}{Jiaying Peng}, \bibinfo{person}{Tao Xie},
  \bibinfo{person}{Zengxu Cao}, \bibinfo{person}{Kun Xu},
  \bibinfo{person}{Xiangnan He}, {and} \bibinfo{person}{Zibin Zheng}.}
  \bibinfo{year}{2020}\natexlab{}.
\newblock \showarticletitle{A Survey of Adversarial Learning on Graphs}.
\newblock \bibinfo{journal}{\emph{CoRR}} (\bibinfo{year}{2020}).
\newblock


\bibitem[Chen et~al\mbox{.}(2022)]%
        {chen2022and}
\bibfield{author}{\bibinfo{person}{Weilong Chen}, \bibinfo{person}{Chenghao
  Huang}, \bibinfo{person}{Weimin Yuan}, \bibinfo{person}{Xiaolu Chen},
  \bibinfo{person}{Wenhao Hu}, \bibinfo{person}{Xinran Zhang}, {and}
  \bibinfo{person}{Yanru Zhang}.} \bibinfo{year}{2022}\natexlab{}.
\newblock \showarticletitle{Title-and-Tag Contrastive Vision-and-Language
  Transformer for Social Media Popularity Prediction}. In
  \bibinfo{booktitle}{\emph{ACM MM}}. \bibinfo{pages}{7008--7012}.
\newblock


\bibitem[Cresci et~al\mbox{.}(2015)]%
        {cresci2015fame}
\bibfield{author}{\bibinfo{person}{Stefano Cresci}, \bibinfo{person}{Roberto
  Di~Pietro}, \bibinfo{person}{Marinella Petrocchi}, \bibinfo{person}{Angelo
  Spognardi}, {and} \bibinfo{person}{Maurizio Tesconi}.}
  \bibinfo{year}{2015}\natexlab{}.
\newblock \showarticletitle{Fame for sale: Efficient detection of fake Twitter
  followers}.
\newblock \bibinfo{journal}{\emph{Decision Support Systems}}
  \bibinfo{volume}{80} (\bibinfo{year}{2015}), \bibinfo{pages}{56--71}.
\newblock


\bibitem[Dai et~al\mbox{.}(2018)]%
        {dai2018adversarial}
\bibfield{author}{\bibinfo{person}{Hanjun Dai}, \bibinfo{person}{Hui Li},
  \bibinfo{person}{Tian Tian}, \bibinfo{person}{Xin Huang},
  \bibinfo{person}{Lin Wang}, \bibinfo{person}{Jun Zhu}, {and}
  \bibinfo{person}{Le Song}.} \bibinfo{year}{2018}\natexlab{}.
\newblock \showarticletitle{Adversarial Attack on Graph Structured Data}. In
  \bibinfo{booktitle}{\emph{ICML}}, Vol.~\bibinfo{volume}{80}.
  \bibinfo{pages}{1123--1132}.
\newblock


\bibitem[Deb et~al\mbox{.}(2019)]%
        {deb2019perils}
\bibfield{author}{\bibinfo{person}{Ashok Deb}, \bibinfo{person}{Luca Luceri},
  \bibinfo{person}{Adam Badawy}, {and} \bibinfo{person}{Emilio Ferrara}.}
  \bibinfo{year}{2019}\natexlab{}.
\newblock \showarticletitle{Perils and Challenges of Social Media and Election
  Manipulation Analysis: The 2018 {US} Midterms}. In
  \bibinfo{booktitle}{\emph{WWW}}. \bibinfo{pages}{237--247}.
\newblock


\bibitem[Dehghan et~al\mbox{.}(2022)]%
        {dehghan2022detecting}
\bibfield{author}{\bibinfo{person}{Ashkan Dehghan}, \bibinfo{person}{Kinga
  Siuta}, \bibinfo{person}{Agata Skorupka}, \bibinfo{person}{Akshat Dubey},
  \bibinfo{person}{Andrei Betlen}, \bibinfo{person}{David Miller},
  \bibinfo{person}{Wei Xu}, \bibinfo{person}{Bogumil Kaminski}, {and}
  \bibinfo{person}{Pawel Pralat}.} \bibinfo{year}{2022}\natexlab{}.
\newblock \showarticletitle{Detecting Bots in Social-networks using Node and
  Structural Embeddings}. In \bibinfo{booktitle}{\emph{DATA}}.
  \bibinfo{pages}{50--61}.
\newblock


\bibitem[Feng et~al\mbox{.}(2022)]%
        {feng2022twibot}
\bibfield{author}{\bibinfo{person}{Shangbin Feng}, \bibinfo{person}{Zhaoxuan
  Tan}, \bibinfo{person}{Herun Wan}, \bibinfo{person}{Ningnan Wang},
  \bibinfo{person}{Zilong Chen}, \bibinfo{person}{Binchi Zhang},
  \bibinfo{person}{Qinghua Zheng}, \bibinfo{person}{Wenqian Zhang},
  \bibinfo{person}{Zhenyu Lei}, \bibinfo{person}{Shujie Yang},
  \bibinfo{person}{Xinshun Feng}, \bibinfo{person}{Qingyue Zhang},
  \bibinfo{person}{Hongrui Wang}, \bibinfo{person}{Yuhan Liu},
  \bibinfo{person}{Yuyang Bai}, \bibinfo{person}{Heng Wang},
  \bibinfo{person}{Zijian Cai}, \bibinfo{person}{Yanbo Wang},
  \bibinfo{person}{Lijing Zheng}, \bibinfo{person}{Zihan Ma},
  \bibinfo{person}{Jundong Li}, {and} \bibinfo{person}{Minnan Luo}.}
  \bibinfo{year}{2022}\natexlab{}.
\newblock \showarticletitle{TwiBot-22: Towards Graph-Based Twitter Bot
  Detection}.
\newblock \bibinfo{journal}{\emph{CoRR}} (\bibinfo{year}{2022}).
\newblock


\bibitem[Feng et~al\mbox{.}(2021b)]%
        {feng2021satar}
\bibfield{author}{\bibinfo{person}{Shangbin Feng}, \bibinfo{person}{Herun Wan},
  \bibinfo{person}{Ningnan Wang}, \bibinfo{person}{Jundong Li}, {and}
  \bibinfo{person}{Minnan Luo}.} \bibinfo{year}{2021}\natexlab{b}.
\newblock \showarticletitle{{SATAR:} {A} Self-supervised Approach to Twitter
  Account Representation Learning and its Application in Bot Detection}. In
  \bibinfo{booktitle}{\emph{CIKM}}. \bibinfo{pages}{3808--3817}.
\newblock


\bibitem[Feng et~al\mbox{.}(2021a)]%
        {feng2021botrgcn}
\bibfield{author}{\bibinfo{person}{Shangbin Feng}, \bibinfo{person}{Herun Wan},
  \bibinfo{person}{Ningnan Wang}, {and} \bibinfo{person}{Minnan Luo}.}
  \bibinfo{year}{2021}\natexlab{a}.
\newblock \showarticletitle{BotRGCN: Twitter bot detection with relational
  graph convolutional networks}. In \bibinfo{booktitle}{\emph{ASONAM}}.
  \bibinfo{pages}{236--239}.
\newblock


\bibitem[Ferrara(2017)]%
        {ferrara2017disinformation}
\bibfield{author}{\bibinfo{person}{Emilio Ferrara}.}
  \bibinfo{year}{2017}\natexlab{}.
\newblock \showarticletitle{Disinformation and social bot operations in the run
  up to the 2017 French presidential election}.
\newblock \bibinfo{journal}{\emph{First Monday}} \bibinfo{volume}{22},
  \bibinfo{number}{8} (\bibinfo{year}{2017}).
\newblock


\bibitem[Ferrara(2020)]%
        {ferrara2020types}
\bibfield{author}{\bibinfo{person}{Emilio Ferrara}.}
  \bibinfo{year}{2020}\natexlab{}.
\newblock \showarticletitle{What types of {COVID-19} conspiracies are populated
  by Twitter bots?}
\newblock \bibinfo{journal}{\emph{First Monday}} \bibinfo{volume}{25},
  \bibinfo{number}{6} (\bibinfo{year}{2020}).
\newblock


\bibitem[Hu et~al\mbox{.}(2023)]%
        {hu2023hyperattack}
\bibfield{author}{\bibinfo{person}{Chao Hu}, \bibinfo{person}{Ruishi Yu},
  \bibinfo{person}{Binqi Zeng}, \bibinfo{person}{Yu Zhan},
  \bibinfo{person}{Ying Fu}, \bibinfo{person}{Quan Zhang},
  \bibinfo{person}{Rongkai Liu}, {and} \bibinfo{person}{Heyuan Shi}.}
  \bibinfo{year}{2023}\natexlab{}.
\newblock \showarticletitle{HyperAttack: Multi-Gradient-Guided White-box
  Adversarial Structure Attack of Hypergraph Neural Networks}.
\newblock \bibinfo{journal}{\emph{CoRR}} (\bibinfo{year}{2023}).
\newblock


\bibitem[Hu et~al\mbox{.}(2020)]%
        {hu2020heterogeneous}
\bibfield{author}{\bibinfo{person}{Ziniu Hu}, \bibinfo{person}{Yuxiao Dong},
  \bibinfo{person}{Kuansan Wang}, {and} \bibinfo{person}{Yizhou Sun}.}
  \bibinfo{year}{2020}\natexlab{}.
\newblock \showarticletitle{Heterogeneous Graph Transformer}. In
  \bibinfo{booktitle}{\emph{WWW}}. \bibinfo{pages}{2704--2710}.
\newblock


\bibitem[Jia et~al\mbox{.}(2020)]%
        {jia2020certified}
\bibfield{author}{\bibinfo{person}{Jinyuan Jia}, \bibinfo{person}{Binghui
  Wang}, \bibinfo{person}{Xiaoyu Cao}, {and} \bibinfo{person}{Neil~Zhenqiang
  Gong}.} \bibinfo{year}{2020}\natexlab{}.
\newblock \showarticletitle{Certified Robustness of Community Detection against
  Adversarial Structural Perturbation via Randomized Smoothing}. In
  \bibinfo{booktitle}{\emph{WWW}}. \bibinfo{pages}{2718--2724}.
\newblock


\bibitem[Kantartopoulos et~al\mbox{.}(2020)]%
        {kantartopoulos2020exploring}
\bibfield{author}{\bibinfo{person}{Panagiotis Kantartopoulos},
  \bibinfo{person}{Nikolaos Pitropakis}, \bibinfo{person}{Alexios Mylonas},
  {and} \bibinfo{person}{Nicolas Kylilis}.} \bibinfo{year}{2020}\natexlab{}.
\newblock \showarticletitle{Exploring adversarial attacks and defences for fake
  twitter account detection}.
\newblock \bibinfo{journal}{\emph{Technologies}} \bibinfo{volume}{8},
  \bibinfo{number}{4} (\bibinfo{year}{2020}), \bibinfo{pages}{64}.
\newblock


\bibitem[Keller et~al\mbox{.}(2020)]%
        {keller2020political}
\bibfield{author}{\bibinfo{person}{Franziska~B Keller}, \bibinfo{person}{David
  Schoch}, \bibinfo{person}{Sebastian Stier}, {and} \bibinfo{person}{JungHwan
  Yang}.} \bibinfo{year}{2020}\natexlab{}.
\newblock \showarticletitle{Political astroturfing on Twitter: How to
  coordinate a disinformation campaign}.
\newblock \bibinfo{journal}{\emph{Political communication}}
  \bibinfo{volume}{37}, \bibinfo{number}{2} (\bibinfo{year}{2020}),
  \bibinfo{pages}{256--280}.
\newblock


\bibitem[Kipf and Welling(2017)]%
        {kipf2016semi}
\bibfield{author}{\bibinfo{person}{Thomas~N. Kipf} {and} \bibinfo{person}{Max
  Welling}.} \bibinfo{year}{2017}\natexlab{}.
\newblock \showarticletitle{Semi-Supervised Classification with Graph
  Convolutional Networks}. In \bibinfo{booktitle}{\emph{ICLR}}.
\newblock


\bibitem[Knauth(2019)]%
        {knauth2019language}
\bibfield{author}{\bibinfo{person}{J{\"{u}}rgen Knauth}.}
  \bibinfo{year}{2019}\natexlab{}.
\newblock \showarticletitle{Language-Agnostic Twitter-Bot Detection}. In
  \bibinfo{booktitle}{\emph{RANLP}}. \bibinfo{pages}{550--558}.
\newblock


\bibitem[Kudugunta and Ferrara(2018)]%
        {kudugunta2018deep}
\bibfield{author}{\bibinfo{person}{Sneha Kudugunta} {and}
  \bibinfo{person}{Emilio Ferrara}.} \bibinfo{year}{2018}\natexlab{}.
\newblock \showarticletitle{Deep neural networks for bot detection}.
\newblock \bibinfo{journal}{\emph{Information Sciences}}  \bibinfo{volume}{467}
  (\bibinfo{year}{2018}), \bibinfo{pages}{312--322}.
\newblock


\bibitem[Liu et~al\mbox{.}(2019)]%
        {liu2019roberta}
\bibfield{author}{\bibinfo{person}{Yinhan Liu}, \bibinfo{person}{Myle Ott},
  \bibinfo{person}{Naman Goyal}, \bibinfo{person}{Jingfei Du},
  \bibinfo{person}{Mandar Joshi}, \bibinfo{person}{Danqi Chen},
  \bibinfo{person}{Omer Levy}, \bibinfo{person}{Mike Lewis},
  \bibinfo{person}{Luke Zettlemoyer}, {and} \bibinfo{person}{Veselin
  Stoyanov}.} \bibinfo{year}{2019}\natexlab{}.
\newblock \showarticletitle{RoBERTa: {A} Robustly Optimized {BERT} Pretraining
  Approach}.
\newblock \bibinfo{journal}{\emph{CoRR}} (\bibinfo{year}{2019}).
\newblock


\bibitem[Lv et~al\mbox{.}(2021)]%
        {lv2021we}
\bibfield{author}{\bibinfo{person}{Qingsong Lv}, \bibinfo{person}{Ming Ding},
  \bibinfo{person}{Qiang Liu}, \bibinfo{person}{Yuxiang Chen},
  \bibinfo{person}{Wenzheng Feng}, \bibinfo{person}{Siming He},
  \bibinfo{person}{Chang Zhou}, \bibinfo{person}{Jianguo Jiang},
  \bibinfo{person}{Yuxiao Dong}, {and} \bibinfo{person}{Jie Tang}.}
  \bibinfo{year}{2021}\natexlab{}.
\newblock \showarticletitle{Are we really making much progress?: Revisiting,
  benchmarking and refining heterogeneous graph neural networks}. In
  \bibinfo{booktitle}{\emph{KDD}}. \bibinfo{pages}{1150--1160}.
\newblock


\bibitem[Ma et~al\mbox{.}(2020)]%
        {ma2020towards}
\bibfield{author}{\bibinfo{person}{Jiaqi Ma}, \bibinfo{person}{Shuangrui Ding},
  {and} \bibinfo{person}{Qiaozhu Mei}.} \bibinfo{year}{2020}\natexlab{}.
\newblock \showarticletitle{Towards More Practical Adversarial Attacks on Graph
  Neural Networks}. In \bibinfo{booktitle}{\emph{NeurIPS}}.
\newblock


\bibitem[Ma et~al\mbox{.}(2021)]%
        {ma2021graph}
\bibfield{author}{\bibinfo{person}{Yao Ma}, \bibinfo{person}{Suhang Wang},
  \bibinfo{person}{Tyler Derr}, \bibinfo{person}{Lingfei Wu}, {and}
  \bibinfo{person}{Jiliang Tang}.} \bibinfo{year}{2021}\natexlab{}.
\newblock \showarticletitle{Graph Adversarial Attack via Rewiring}. In
  \bibinfo{booktitle}{\emph{KDD}}. \bibinfo{pages}{1161--1169}.
\newblock


\bibitem[Pham et~al\mbox{.}(2022)]%
        {pham2022bot2vec}
\bibfield{author}{\bibinfo{person}{Phu Pham}, \bibinfo{person}{Loan~TT Nguyen},
  \bibinfo{person}{Bay Vo}, {and} \bibinfo{person}{Unil Yun}.}
  \bibinfo{year}{2022}\natexlab{}.
\newblock \showarticletitle{Bot2Vec: A general approach of intra-community
  oriented representation learning for bot detection in different types of
  social networks}.
\newblock \bibinfo{journal}{\emph{Information Systems}}  \bibinfo{volume}{103}
  (\bibinfo{year}{2022}), \bibinfo{pages}{101771}.
\newblock


\bibitem[Rieck et~al\mbox{.}(2019)]%
        {rieck2019persistent}
\bibfield{author}{\bibinfo{person}{Bastian Rieck}, \bibinfo{person}{Christian
  Bock}, {and} \bibinfo{person}{Karsten~M. Borgwardt}.}
  \bibinfo{year}{2019}\natexlab{}.
\newblock \showarticletitle{A Persistent Weisfeiler-Lehman Procedure for Graph
  Classification}. In \bibinfo{booktitle}{\emph{ICML}},
  Vol.~\bibinfo{volume}{97}. \bibinfo{pages}{5448--5458}.
\newblock


\bibitem[Sang et~al\mbox{.}(2022)]%
        {sang2022benign}
\bibfield{author}{\bibinfo{person}{Jitao Sang}, \bibinfo{person}{Xian Zhao},
  \bibinfo{person}{Jiaming Zhang}, {and} \bibinfo{person}{Zhiyu Lin}.}
  \bibinfo{year}{2022}\natexlab{}.
\newblock \showarticletitle{Benign Adversarial Attack: Tricking Models for
  Goodness}. In \bibinfo{booktitle}{\emph{ACM MM}}.
  \bibinfo{pages}{6883--6889}.
\newblock


\bibitem[Schlichtkrull et~al\mbox{.}(2018)]%
        {schlichtkrull2018modeling}
\bibfield{author}{\bibinfo{person}{Michael~Sejr Schlichtkrull},
  \bibinfo{person}{Thomas~N. Kipf}, \bibinfo{person}{Peter Bloem},
  \bibinfo{person}{Rianne van~den Berg}, \bibinfo{person}{Ivan Titov}, {and}
  \bibinfo{person}{Max Welling}.} \bibinfo{year}{2018}\natexlab{}.
\newblock \showarticletitle{Modeling Relational Data with Graph Convolutional
  Networks}. In \bibinfo{booktitle}{\emph{ESWC}}, Vol.~\bibinfo{volume}{10843}.
  \bibinfo{pages}{593--607}.
\newblock


\bibitem[Sun et~al\mbox{.}(2020)]%
        {sun2020adversarial}
\bibfield{author}{\bibinfo{person}{Yiwei Sun}, \bibinfo{person}{Suhang Wang},
  \bibinfo{person}{Xianfeng Tang}, \bibinfo{person}{Tsung{-}Yu Hsieh}, {and}
  \bibinfo{person}{Vasant~G. Honavar}.} \bibinfo{year}{2020}\natexlab{}.
\newblock \showarticletitle{Adversarial Attacks on Graph Neural Networks via
  Node Injections: {A} Hierarchical Reinforcement Learning Approach}. In
  \bibinfo{booktitle}{\emph{WWW}}. \bibinfo{pages}{673--683}.
\newblock


\bibitem[Tan et~al\mbox{.}(2022)]%
        {tan2022efficient}
\bibfield{author}{\bibinfo{person}{Yunpeng Tan}, \bibinfo{person}{Fangyu Liu},
  \bibinfo{person}{Bowei Li}, \bibinfo{person}{Zheng Zhang}, {and}
  \bibinfo{person}{Bo Zhang}.} \bibinfo{year}{2022}\natexlab{}.
\newblock \showarticletitle{An Efficient Multi-View Multimodal Data Processing
  Framework for Social Media Popularity Prediction}. In
  \bibinfo{booktitle}{\emph{ACM MM}}. \bibinfo{pages}{7200--7204}.
\newblock


\bibitem[Tao et~al\mbox{.}(2021)]%
        {tao2021single}
\bibfield{author}{\bibinfo{person}{Shuchang Tao}, \bibinfo{person}{Qi Cao},
  \bibinfo{person}{Huawei Shen}, \bibinfo{person}{Junjie Huang},
  \bibinfo{person}{Yunfan Wu}, {and} \bibinfo{person}{Xueqi Cheng}.}
  \bibinfo{year}{2021}\natexlab{}.
\newblock \showarticletitle{Single Node Injection Attack against Graph Neural
  Networks}. In \bibinfo{booktitle}{\emph{CIKM}}. \bibinfo{pages}{1794--1803}.
\newblock


\bibitem[Torusda{\u{g}} et~al\mbox{.}(2020)]%
        {torusdaug2020we}
\bibfield{author}{\bibinfo{person}{M~Bu{\u{g}}ra Torusda{\u{g}}},
  \bibinfo{person}{Mucahid Kutlu}, {and} \bibinfo{person}{Ali~Ayd{\i}n
  Sel{\c{c}}uk}.} \bibinfo{year}{2020}\natexlab{}.
\newblock \showarticletitle{Are we secure from bots? Investigating
  vulnerabilities of botometer}. In \bibinfo{booktitle}{\emph{UBMK}}.
  \bibinfo{pages}{343--348}.
\newblock


\bibitem[von~der Weth et~al\mbox{.}(2020)]%
        {von2020helping}
\bibfield{author}{\bibinfo{person}{Christian von~der Weth},
  \bibinfo{person}{Ashraf~M. Abdul}, \bibinfo{person}{Shaojing Fan}, {and}
  \bibinfo{person}{Mohan~S. Kankanhalli}.} \bibinfo{year}{2020}\natexlab{}.
\newblock \showarticletitle{Helping Users Tackle Algorithmic Threats on Social
  Media: {A} Multimedia Research Agenda}. In \bibinfo{booktitle}{\emph{ACM
  MM}}. \bibinfo{pages}{4425--4434}.
\newblock


\bibitem[Wang et~al\mbox{.}(2020)]%
        {wang2020mgaattack}
\bibfield{author}{\bibinfo{person}{Lina Wang}, \bibinfo{person}{Kang Yang},
  \bibinfo{person}{Wenqi Wang}, \bibinfo{person}{Run Wang}, {and}
  \bibinfo{person}{Aoshuang Ye}.} \bibinfo{year}{2020}\natexlab{}.
\newblock \showarticletitle{MGAAttack: Toward More Query-efficient Black-box
  Attack by Microbial Genetic Algorithm}. In \bibinfo{booktitle}{\emph{ACM
  MM}}. \bibinfo{pages}{2229--2236}.
\newblock


\bibitem[Xu et~al\mbox{.}(2019b)]%
        {xu2020graph}
\bibfield{author}{\bibinfo{person}{Bingbing Xu}, \bibinfo{person}{Huawei Shen},
  \bibinfo{person}{Qi Cao}, \bibinfo{person}{Keting Cen}, {and}
  \bibinfo{person}{Xueqi Cheng}.} \bibinfo{year}{2019}\natexlab{b}.
\newblock \showarticletitle{Graph Convolutional Networks using Heat Kernel for
  Semi-supervised Learning}. In \bibinfo{booktitle}{\emph{IJCAI}}.
  \bibinfo{pages}{1928--1934}.
\newblock


\bibitem[Xu et~al\mbox{.}(2020b)]%
        {xu2020adversarial}
\bibfield{author}{\bibinfo{person}{Han Xu}, \bibinfo{person}{Yao Ma},
  \bibinfo{person}{Haochen Liu}, \bibinfo{person}{Debayan Deb},
  \bibinfo{person}{Hui Liu}, \bibinfo{person}{Jiliang Tang}, {and}
  \bibinfo{person}{Anil~K. Jain}.} \bibinfo{year}{2020}\natexlab{b}.
\newblock \showarticletitle{Adversarial Attacks and Defenses in Images, Graphs
  and Text: {A} Review}.
\newblock \bibinfo{journal}{\emph{Int. J. Autom. Comput.}}
  \bibinfo{volume}{17}, \bibinfo{number}{2} (\bibinfo{year}{2020}),
  \bibinfo{pages}{151--178}.
\newblock


\bibitem[Xu et~al\mbox{.}(2019a)]%
        {xu2019topology}
\bibfield{author}{\bibinfo{person}{Kaidi Xu}, \bibinfo{person}{Hongge Chen},
  \bibinfo{person}{Sijia Liu}, \bibinfo{person}{Pin{-}Yu Chen},
  \bibinfo{person}{Tsui{-}Wei Weng}, \bibinfo{person}{Mingyi Hong}, {and}
  \bibinfo{person}{Xue Lin}.} \bibinfo{year}{2019}\natexlab{a}.
\newblock \showarticletitle{Topology Attack and Defense for Graph Neural
  Networks: An Optimization Perspective}. In \bibinfo{booktitle}{\emph{IJCAI}}.
  \bibinfo{pages}{3961--3967}.
\newblock


\bibitem[Xu et~al\mbox{.}(2021)]%
        {xu2020towards}
\bibfield{author}{\bibinfo{person}{Qiuling Xu}, \bibinfo{person}{Guanhong Tao},
  \bibinfo{person}{Siyuan Cheng}, {and} \bibinfo{person}{Xiangyu Zhang}.}
  \bibinfo{year}{2021}\natexlab{}.
\newblock \showarticletitle{Towards Feature Space Adversarial Attack by Style
  Perturbation}. In \bibinfo{booktitle}{\emph{AAAI}}.
  \bibinfo{pages}{10523--10531}.
\newblock


\bibitem[Xu et~al\mbox{.}(2020a)]%
        {xu2020learning}
\bibfield{author}{\bibinfo{person}{Xing Xu}, \bibinfo{person}{Jiefu Chen},
  \bibinfo{person}{Jinhui Xiao}, \bibinfo{person}{Zheng Wang},
  \bibinfo{person}{Yang Yang}, {and} \bibinfo{person}{Heng~Tao Shen}.}
  \bibinfo{year}{2020}\natexlab{a}.
\newblock \showarticletitle{Learning Optimization-based Adversarial
  Perturbations for Attacking Sequential Recognition Models}. In
  \bibinfo{booktitle}{\emph{ACM MM}}. \bibinfo{pages}{2802--2822}.
\newblock


\bibitem[Yang et~al\mbox{.}(2021)]%
        {yang2021model}
\bibfield{author}{\bibinfo{person}{Xiao Yang}, \bibinfo{person}{Yinpeng Dong},
  \bibinfo{person}{Wenzhao Xiang}, \bibinfo{person}{Tianyu Pang},
  \bibinfo{person}{Hang Su}, {and} \bibinfo{person}{Jun Zhu}.}
  \bibinfo{year}{2021}\natexlab{}.
\newblock \showarticletitle{Model-Agnostic Meta-Attack: Towards Reliable
  Evaluation of Adversarial Robustness}.
\newblock \bibinfo{journal}{\emph{CoRR}} (\bibinfo{year}{2021}).
\newblock


\bibitem[Yuan et~al\mbox{.}(2021)]%
        {yuan2021meta}
\bibfield{author}{\bibinfo{person}{Zheng Yuan}, \bibinfo{person}{Jie Zhang},
  \bibinfo{person}{Yunpei Jia}, \bibinfo{person}{Chuanqi Tan},
  \bibinfo{person}{Tao Xue}, {and} \bibinfo{person}{Shiguang Shan}.}
  \bibinfo{year}{2021}\natexlab{}.
\newblock \showarticletitle{Meta Gradient Adversarial Attack}. In
  \bibinfo{booktitle}{\emph{ICCV}}. \bibinfo{pages}{7728--7737}.
\newblock


\bibitem[Z{\"{u}}gner et~al\mbox{.}(2019)]%
        {zugner2018adversarial}
\bibfield{author}{\bibinfo{person}{Daniel Z{\"{u}}gner}, \bibinfo{person}{Amir
  Akbarnejad}, {and} \bibinfo{person}{Stephan G{\"{u}}nnemann}.}
  \bibinfo{year}{2019}\natexlab{}.
\newblock \showarticletitle{Adversarial Attacks on Neural Networks for Graph
  Data}. In \bibinfo{booktitle}{\emph{IJCAI}}. \bibinfo{pages}{6246--6250}.
\newblock


\end{thebibliography}

\appendix

\section{Alternative constraints}

In Sec.~\ref{constraint_settings}, we set the followers count to zero, which is to minimize the costs associated with buying followers, making the attack easier to carry out. However, it should be noted that setting the followers count to zero does not mean that it must be zero. According to \cite{feng2021satar} showing that the number of followers is positively correlated with the probability that a node is detected as human, if the injected node can successfully attack without followers, then it can successfully attack with any number of followers. Thus, we relax the restriction on zero follower count and conduct more experiments. In the following experiments, we set the follower count as 600 and reduce the status to 18,000 for TwiBot-22. The rest of the settings are the same as Table~\ref{constraint_22}. 

\begin{table}[htpb]
  \centering
  \caption{Analysis of followers count on TwiBot-22 (\%)}
  \label{followers_status}
  \begin{tabular}{ccc}
    \hline
    Method      & attack success rate  & new node become bot  \\
    \hline
    GCN   & 93.27 ± 1.38   & 10.31 ± 11.28  \\
    HGT   & 89.00 ± 4.00   & 5.62 ± 11.24  \\
    Simple-HGN   & 74.37 ± 2.70   & 18.14 ± 30.44   \\
    R-GCN     & 76.53 ± 2.40   & 3.34 ± 6.05   \\
    \hline
\end{tabular}
\end{table}

As shown in Table~\ref{followers_status}, increasing the followers count still keeps the success rate of the attacks high. At the same time, increasing the followers count can significantly reduce the status requirements of the injection nodes. This shows that attackers can set the constraints on their own according to the actual situation.

\end{document}